\documentclass[3p,times,twocolumn,procedia]{elsarticle}

\usepackage{ecrc}

\volume{00}

\firstpage{1}

\journalname{Nuclear Physics B Proceedings Supplement}

\runauth{}

\jid{nuphbp}

\jnltitlelogo{Nuclear Physics B Proceedings Supplement}

\CopyrightLine{2011}{Published by Elsevier Ltd.}

\usepackage{amssymb}

\usepackage[figuresright]{rotating}

\begin{document}

\begin{frontmatter}



\dochead{\small{IFJPAN-IV-2014-15}}

\title{Study of the tau meson decay modes with Monte Carlo generator TAUOLA. Status and perspectives.}


\author{Olga Shekhovtsova}
\ead{olga.shekhovtsova@ifj.edu.pl}
\address{Institute of Nuclear Physics PAN ul. Radzikowskiego 152 31-342 Krakow, Poland \\
           Kharkov Institute of Physics and Technology   61108, Akademicheskaya,1, Kharkov, Ukraine}

\begin{abstract}

In the last two years
substantial progress for the simulation of the process: $\tau^- \to  \pi^-\pi^+\pi^- \nu_\tau$
by the Monte Carlo generator TAUOLA
was achieved. It is related to a new parametrization of the corresponding
hadronic current based on the Resonance Chiral Lagrangian and the recent
availability of the unfolded distributions from the BaBar Collaboration analysis for all
invariant hadronic masses. The theoretical model parameters were fitted to the one-dimensional
distributions provided by the BaBar Collaboration and results of the fit are discussed. 
A set of the hadronic currents for other final states with two and three
pseudoscalars is also installed in TAUOLA and the preliminary results for
fitting $K^+K^-\pi^-$ and $\pi^0\pi^-$ to BaBar and Belle data are presented. 
\end{abstract}


\end{frontmatter}


\begin{figure*}
\centering
\vspace{-1.5cm} 
\includegraphics[width=.3\textwidth]{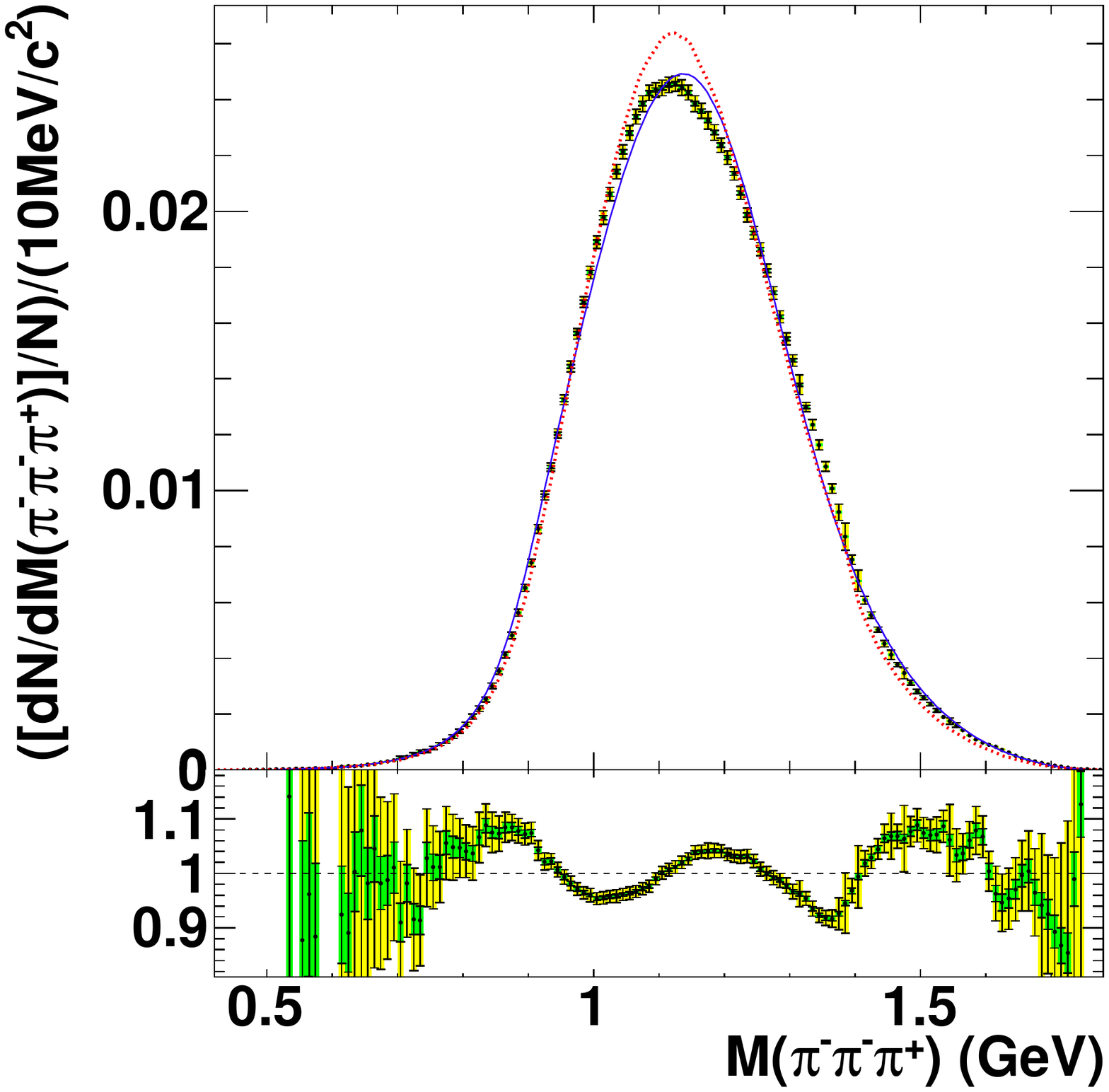}
\includegraphics[width=.30\textwidth]{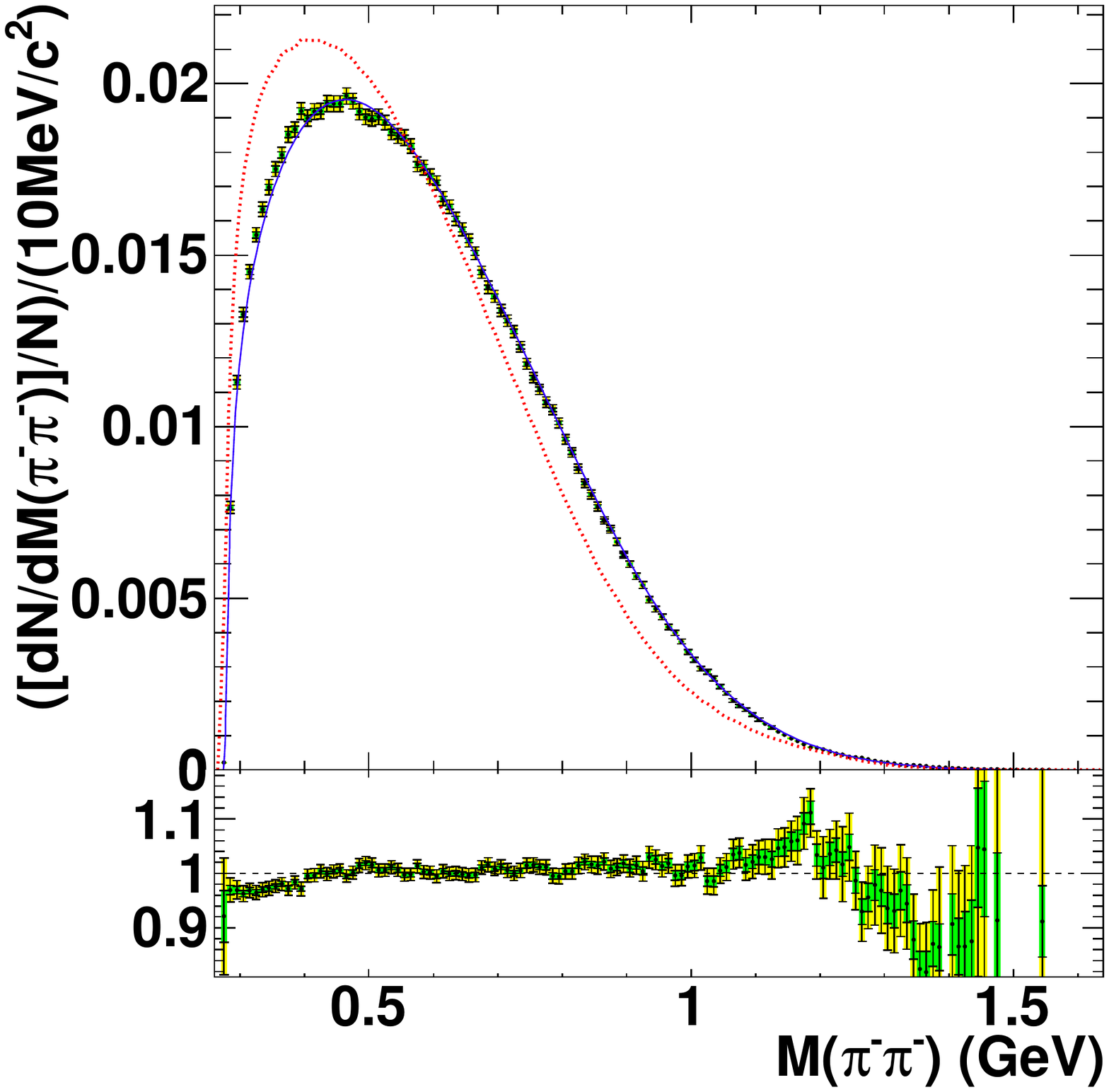}
\includegraphics[width=.30\textwidth]{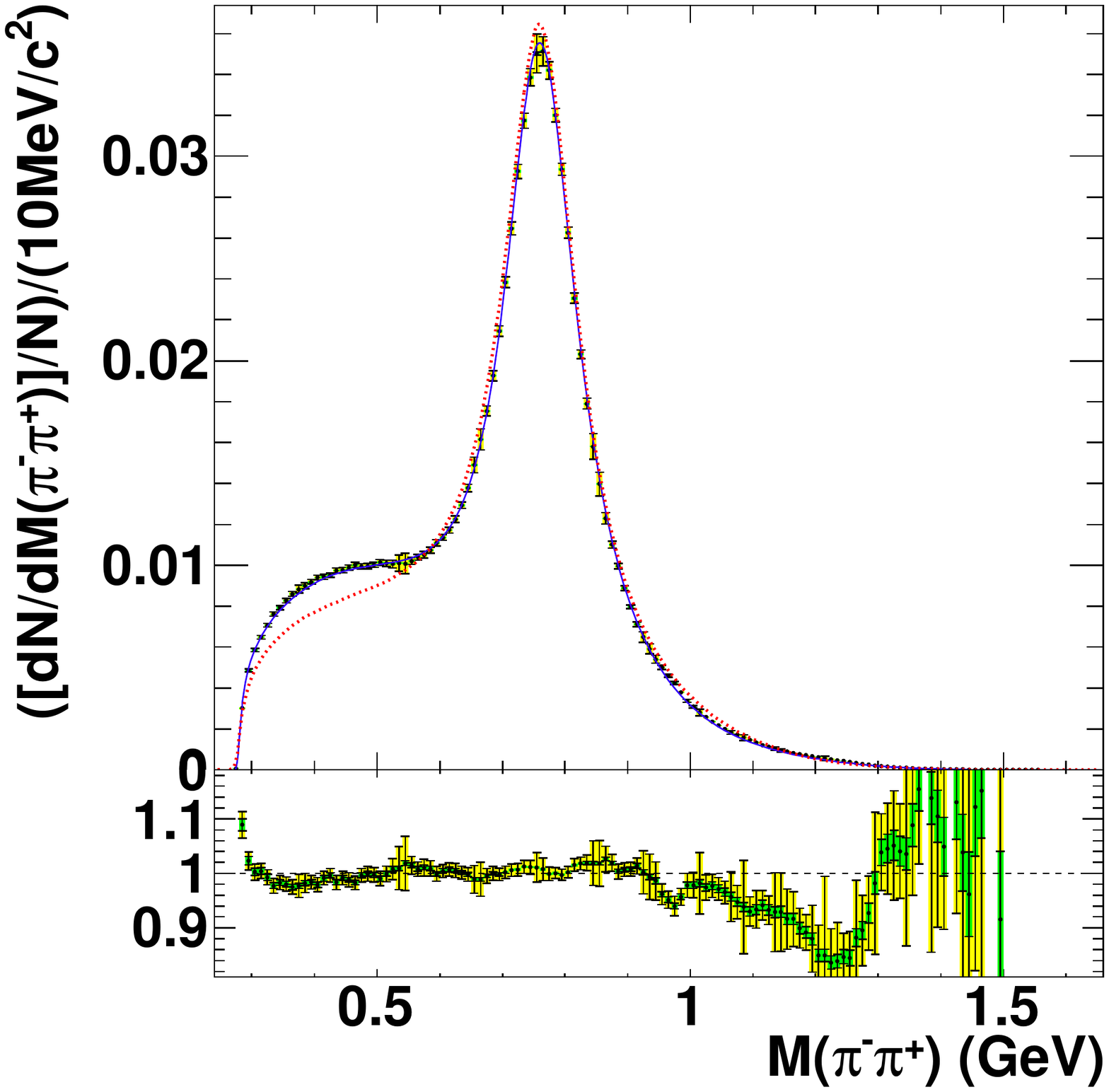}
\vspace{-0.3cm} 
\caption{The $\tau^- \to \pi^- \pi^-\pi^+\nu_\tau$ decay 
invariant mass distribution of the three-pion system (left panel) and two-pion pairs (central and right panels).  The BaBar measurements \cite{Nugent:2013ij} are represented 
by the data points, 
with the results from the R$\chi$L current as described in the text (blue line) and the
 old fit from CLEO as detailed in Refs.~\cite{Davidson:2010rw} (red-dashed line) overlaid.   
At the bottom of the figures the ratio 
of the new  R$\chi$L prediction to the data is given.
The parameters used in our model are collected in Table 1 of \cite{Nugent:2013hxa}. 
\label{fig:res}}
\end{figure*}

\section{Introduction}
\label{sec:introduction}
TAUOLA \cite{Jadach:1993hs} is a Monte Carlo (MC) generator dedicated to the generation of $\tau$-lepton decays and it is
used in the analysis of experimental data both at B-factories, BaBar \cite{Aubert:2007mh} and Belle \cite{Ryu:2014vpc} Collaborations,
and LHC \cite{Davidson:2010rw}. The generator simulates more than twenty decay modes, including both leptonic and hadronic modes.
The leptonic decay modes of the $\tau$ lepton allow to test the universality of the lepton
couplings to the gauge bosons. The hadronic decays (in fact, the $\tau$ lepton due to its high mass is the only one that can decay into
hadrons) give information about the hadronization mechanism and resonance dynamics in the energy region where the methods
of perturbative QCD cannot be applied. Also hadronic flavour- and CP-violating decays of the $\tau$ lepton allow to search
for new physics scenarios. In addition, the tau lepton decay data allows us to measure  the Standard Model parameters, such as the strong coupling constant, the quark-mixing matrix, the strange quark mass etc. 

The main problem in description of the hadronic decay modes of the $\tau$ lepton is the lack of a theory coming from the first principle in the energy region populated by the resonances (i.e. in the region of 1-2 GeV). The hadronic currents implemented in the first version of TAUOLA \cite{Jadach:1993hs} as well as in the internal versions of the code used by BaBar
and Belle are based on Vector Meson Dominance (VMD) approach. As shown in \cite{Bevan:2014iga}, Figs. 20.6.3 and 20.6.4, the current version of
TAUOLA, used by Belle collaboration does not give a satisfactory description of the data. This indicates that the model in the generator code should be updated. 
In last three years we have been working on the partial upgrade of the generator TAUOLA by using the results for the hadronic currents calculated within Resonance Chiral Lagrangian (R$\chi$L)  formalism and fitting the model parameters to the available experimental data from B-factories.
The R$\chi$L approach succeeds in reproducing low energy results, predicted by Chiral Perturbation Theory, 
at least, at the next-to-leading order and also complies with QCD high energy constraints.  Alternative ad-hoc models, as VMD mentioned above, lack a link with QCD and thus can, at most, reproduce the leading order properties.
Presently the R$\chi$L currents for the main two-meson (final states with two pion,
pion-kaon, two kaons) and three-pseudoscalar (three pion, two kaon-one pion) decay modes have been  installed into TAUOLA. This set covers more than 88$\%$ of the hadronic $\tau$ decay width. 
The implementation of the currents, the related technical tests as well as the necessary theoretical concepts  are documented in \cite{Shekhovtsova:2012ra}. 

Studies of the one-dimensional three-prong decay modes by BaBar \cite{Nugent:2013ij} allowed us to compare  R$\chi$L predictions with the measured data.
 We started with the $\pi^-\pi^-\pi^+$ mode. The choice of this channel was motivated by its
relatively large branching ratio
and the already non-trivial dynamics of three-pion final state. 
 The first comparison to the BaBar preliminary data demonstrated 
satisfactory agreement with the three pion invariant mass spectrum and a mismatch in the low energy tail in two pion invariant distributions \cite{Shekhovtsova:2013rb}.
 This mismatch indicates that the lack of the scalar $f_0(500)$ resonance, so called $\sigma$ meson, in the R$\chi$L formalism may be responsible for this discrepency. A modification to the R$\chi$L approach to include the $\sigma$ meson was proposed in  \cite{Nugent:2013hxa} and as a result the agreement with the data was improved by a factor of about eight.

The paper is organized as follows. In Section 2 the theoretical framework for $\pi^-\pi^-\pi^+$ current as well as numerical results 
of the fit to data from BaBar  is presented. Section 3 contains the first results of the fit for $\pi^0\pi^-$ and $ K^+K^-\pi^-$ to data from BaBar and Belle experiments, correspodningly. The summary of Section 4 closes the paper.

\section{Decay mode  $\tau^- \to \pi^-\pi^-\pi^+ \nu_\tau$. Fit to  BaBar data: numerical results and tests}\label{sect:3pion}
For the final state $\pi^-\pi^-\pi^+$ the following mechanisms of  production have been taken into account :
\begin{itemize}
\item double resonance mechanism of production $\tau^- \to a_1^- \nu_\tau \to \pi^-(\rho ; \sigma)   \nu_\tau \to  \pi^-\pi^-\pi^+ \nu_\tau \; ,$
\item  single resonance mechanism of production $\tau^- \to \pi^-(\rho; \sigma) 
 \nu_\tau \to  \pi^-\pi^-\pi^+ \nu_\tau\; ,$
\item  a chiral contribution (a direct decay, without production of any
 intermediate resonance) .
\end{itemize}
The exact form of the hadronic current can be found in \cite{Nugent:2013hxa}, Section II.

A resonance mechanism for the production of the $\pi^+\pi^-$ via
the lightest scalar resonance, $f_0(500)$, not present in the previous version of TAUOLA \cite{Shekhovtsova:2012ra,Shekhovtsova:2013rb}, has been included in the simulation. The nature of the $\sigma$ resonance,  is still unclear and various descriptions are proposed by different groups: meson-meson molecular, tetraquark  etc. \cite{Guo:2013nja}. 
As a result of this behavior the $\sigma$ resonance cannot be easily included in the R$\chi$L formalism. In view of this we have decided to  
incorporate the intermediate $\sigma$ meson state in a form that reproduces the R$\chi$L current structure (i.e. contains single and double resonance contributions, see above), and represent the $\sigma$ meson  by an s-wave Breit-Wigner function following a phenomelogical approximation. 

To obtain the numerical values of the model parameters (the mass of resonances, vertex couplings etc, for details, see  \cite{Nugent:2013hxa}) the three one-dimensional distributions, 
namely $d\Gamma/dm_{\pi^-\pi^-\pi^+}$, $d\Gamma/dm_{\pi^-\pi^-}$ and $d\Gamma/dm_{\pi^-\pi^+}$, have been fitted to the BaBar preliminary data \cite{Nugent:2013ij}.  The partial width of the $\tau^- \to \pi^-\pi^-\pi^+ \nu_\tau$ decay
is normalized to that measured by BaBar $\Gamma= (2.00\pm 0.03\%)\cdot 10^{-13}$ GeV \cite{Aubert:2007mh}. 

The first problem related with the fit is the calculation of the width in the  $a_1(1260)$ propagator. The $a_1$ width can be written down as the imaginary
part of the two-loop axial-vector-axial-vector correlation function with suitable flavour indices and is a double integral
 of the same hadronic form factors that fill in the hadronic currents (for details, see \cite{Shekhovtsova:2012ra}, Section 3).  Calculating this integral at each point in the parameter space would degrade the performance of the MC generator by
a factor of a thousand (taking into account all decays rejected in the
MC process).
To avoid this, TAUOLA uses a 1000-point precalculated table of the resonance width which is later interpolated to obtain precise value for each point in phase space. Our first attempt was to calculate the width only at the starting point of the fitting procedure and not to recalculate it during the fit. However, the lack of
proper recalculation of this resonance turned out to greatly influence the results. A fitting procedure that relies on the width calculated only once
ends up in a minimum completely off the global minimum found when this width
is properly recalculated for each point in the parameter space. Another approximation used in the $a_1$ width calculation was based on an estimate of the $g$-function, as detailed in \cite{Nugent:2013hxa}. Finally, after the introduction of the parallelized calculation, we were able to incorporate the precise calculations of the $a_1$ width table into the project.
\begin{figure}\label{fig:prel2pi}
\vspace{-1.5cm} 
\centering
\includegraphics[width = 0.3\textwidth]{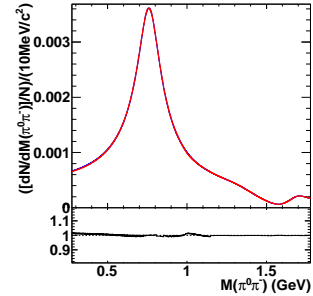}
\vspace{-0.3cm} 
\caption{The $\tau^- \to \pi^0 \pi^-\nu_\tau$ decay 
invariant mass distribution }
\end{figure}

To fit the data we used the MINUIT package through the ROOT framework 
and the fit result is presented in 
Fig.~\ref{fig:res}, for the numerical values of the model parameters see Table 1 in \cite{Nugent:2013hxa}. The goodness of the fit is quantified by  $\chi^2/ndf = 6658/401$. To compare with the previous result \cite{Shekhovtsova:2013rb} we have estimated the $\chi^2$ value using 
the combined statistical 
and systematic uncertainties since only the total covariance matrix was publicly available at that time. For the present results 
we obtain $\chi^2/ndf = 910/401$, when the 
total covariance matrix is used and the conditions enabling direct
comparisons are fulfilled.
 Thus is eight times better than the 
previous result.
 The statistical uncertainties were determined using the {\tt HESSE}
routine from MINUIT
under the assumption 
that the correlations between distributions and the correlations related to having two entries per event in 
the $\pi^{-}\pi^{+}$ distribution can be  neglected. The fit results with estimated systematical and statistical errors, the statistical and the systematic correlation matrices   are collected in tables 3, 4  and 5 of \cite{Nugent:2013hxa}, correspondingly.

The following test has been done to check whether the obtained minimum is a global one and does not depend on the starting parameter values. We start from a random scan of  $2.1*10^5$ points and select 1000 events with the best $\chi^2$, out of which 20 points with maximum distance between then are retained and then these points are used as a start point for the full fit. We find that
more than a half converges to the minimum (Table 1 in  \cite{Nugent:2013hxa}), the rest either reach the limits of the parametric range or converge to local minimum with higher $\chi^2$. Therefore, we conclude that the obtained result is stable and does not depend on the initial value of the fitting parameters. 
As an additional cross check we calculated the partial width resulting from the phase space integration
of the matrix element $\Gamma_{\tau^- \to \pi^- \pi^-  \pi^+ \nu_\tau } = 1.9974 \cdot 10^{-13}$ GeV which
agrees with the one measured by BaBar    $\Gamma_{\tau^- \to \pi^- \pi^-  \pi^+ \nu_\tau } = (2.00\pm 0.03\%)\cdot 10^{-13}$ GeV \cite{Aubert:2007mh}.

In addition, based on the fitted values of the R$\chi$L parameters we estimated the $\pi^0\pi^0\pi^-$ partial width: $\Gamma =  ( 2.1211\pm 0.016\%)\cdot 10^{-13}$ GeV that is $1\%$ higher than the central PDG value and within the errors cited by PDG.

\section{Decay modes  $\tau^- \to \pi^0\pi^-\nu_\tau$ and $\tau^- \to K^+K^-\pi^- \nu_\tau$. Preliminary results}\label{sect:other}

\begin{figure}\label{fig:prelKpiK}
\vspace{-1.2cm} 
\includegraphics[width = 0.23\textwidth]{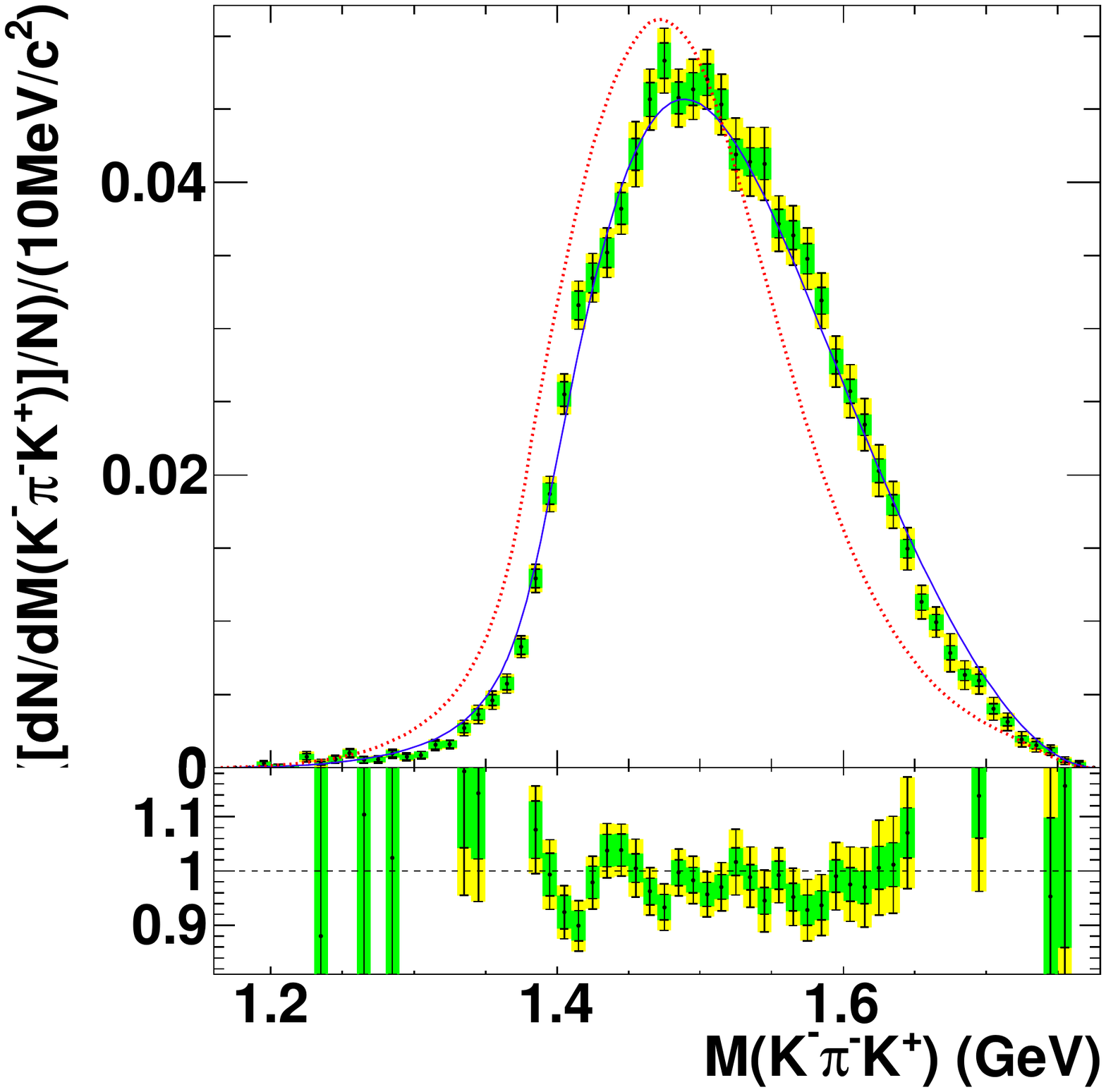}
\includegraphics[width = 0.23\textwidth]{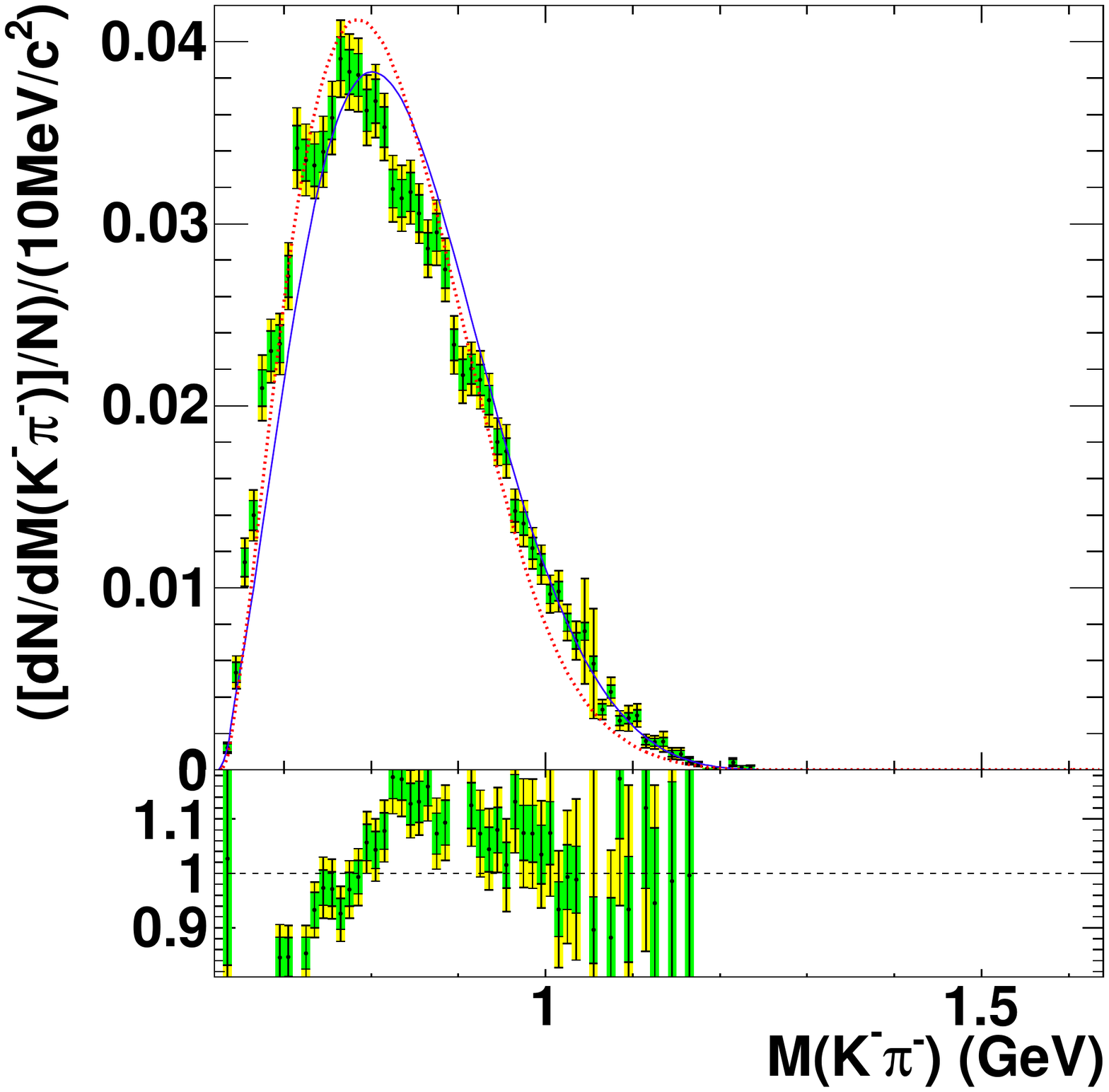}
\vspace{-0.3cm} 
\includegraphics[width = 0.23\textwidth]{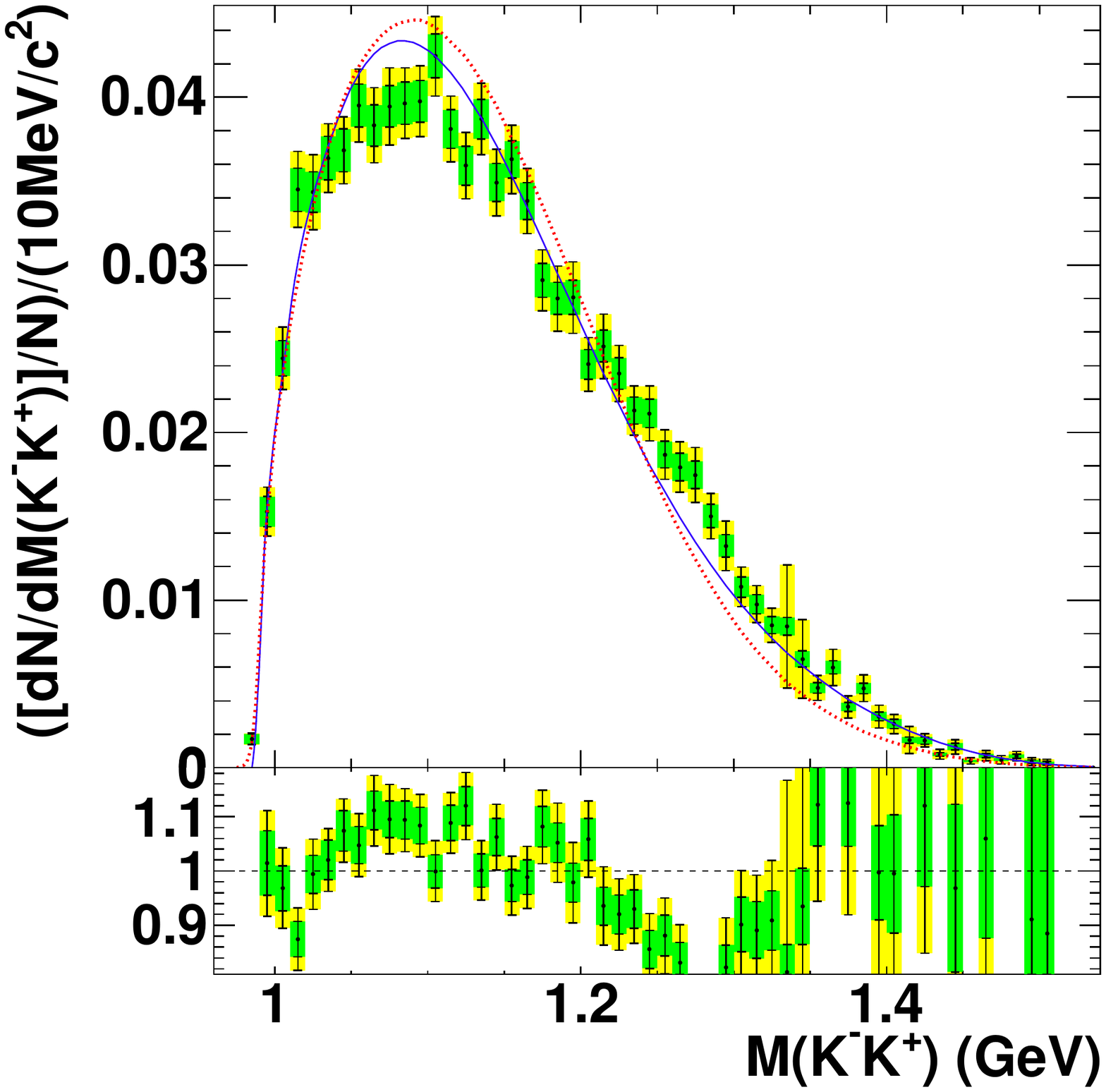}
\includegraphics[width = 0.23\textwidth]{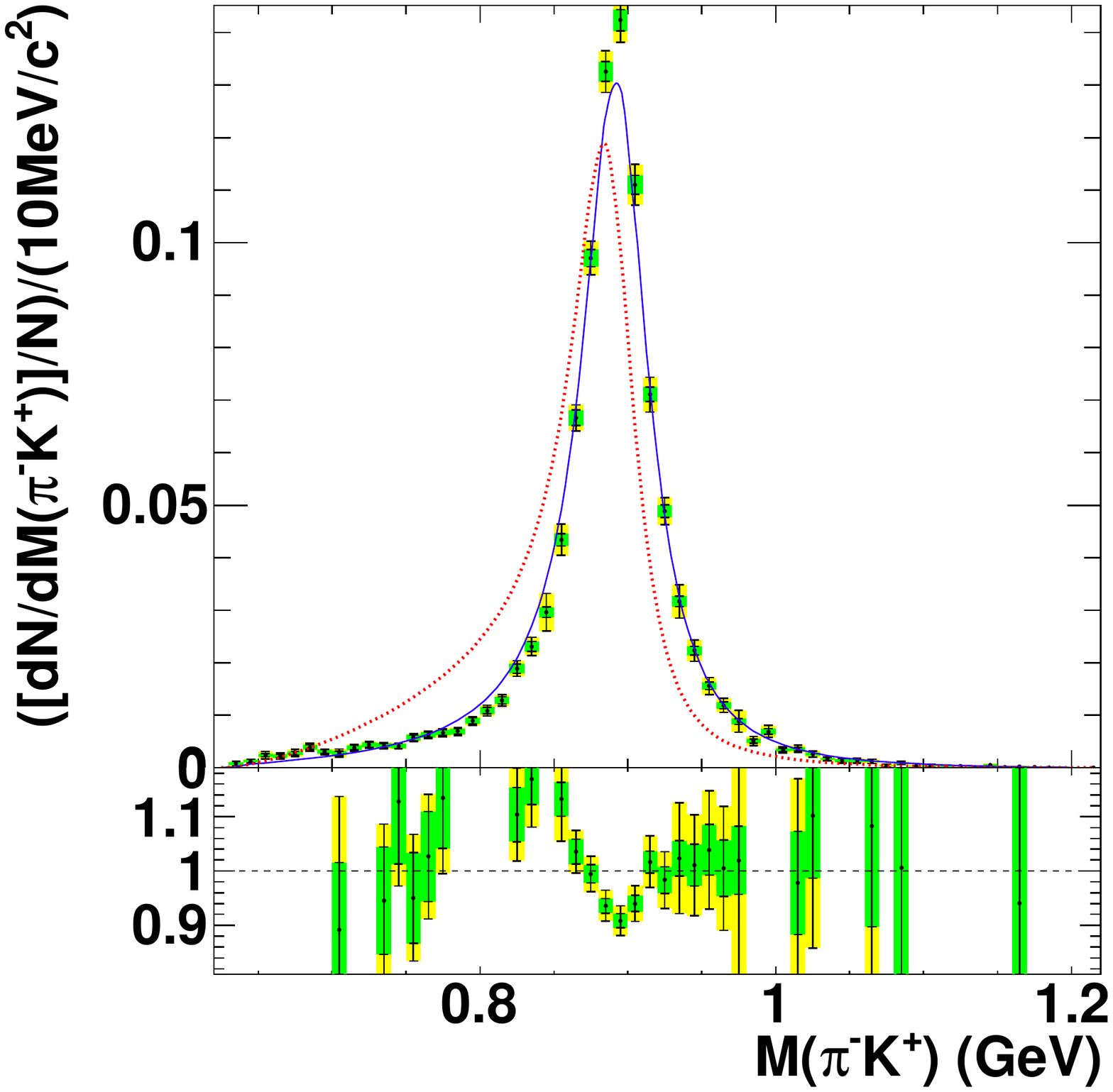}
\caption{The $\tau^- \to K^+ K^-\pi^-\nu_\tau$ decay 
invariant mass distribution of three and two meson system. For the description of the plots see Fig. 1.}
\end{figure}

The first preliminary results for the $\pi^0\pi^-$ and $K^+K^-\pi^-$ modes are presented in Figs.~\ref{fig:prel2pi} and \ref{fig:prelKpiK}.
For the former, we have fitted the absolute value of the pion form factor calculated within the dispersive representation \cite{Dumm:2013zh} and to the  Belle parametrization for the pion form factor, Eqs. (11)-(14) in \cite{Fujikawa:2008ma} (at present the experimental errors are not included in the fit).
The latter was carried out in the generalized version of the fitting strategy used for the $\pi^-\pi^-\pi^+$ mode presented above. In our approach we used the $a_1$ width calculated only at the beginning of the fitting and did not change it during the fit. 
An improved procedure might require a common fit of both $\pi^-\pi^-\pi^+$ and $K^+K^-\pi^-$ modes.   



\section{Conclusion}
In this paper we have discussed the status of the TAUOLA project. The main attention was devoted to the results of the one-dimensional fit for the  $\tau^- \to \pi^-\pi^+\pi^-\nu_\tau$ decay mode to the preliminary BaBar data. The theoretical approach was based on the Resonance Chiral Lagrangian with an additional  modification to the current to include the sigma meson.  As a result, we improved agreement with the data by a factor of 
about eight compared with the previous results \cite{Shekhovtsova:2013rb}. We tested that the obtained result corresponds to a global minimum and that the fitting procedure does not depend on the initial values of the model parameters.

Nonetheless, the model shows discrepancies in the high energy tail of the three pion invariant mass spectrum, that may be related with missing resonances, e.g. $a_1(1640)$, in the corresponding theoretical approach. We will come again to this point in future multidimensional analysis. 

We presented the first results of the generalization of the fitting strategy to the case of an arbitrary three meson tau decay, specializing to the  $K^+K^-\pi^-$ decay mode. In addition we fitted the two pion form factor \cite{Dumm:2013zh} to the Belle parametrization for it. The technical study of the fit stability and the correlation of the parameters is in progress.  

\section{Acknowledgements}
This research was supported in part by Foundation of Polish Science
grant POMOST/2013-7/12, that is co-financed from European Union, Regional
Development
Fund and from funds of Polish National Science
Centre under decisions  DEC-2011/03/B/ST2/00107.

\bibliographystyle{elsarticle-num}



\end{document}